\documentclass[fleqn,usenatbib]{mnras}

\usepackage{newtxtext,newtxmath}
\usepackage{color}
\usepackage[T1]{fontenc}
\usepackage[symbol]{footmisc}
\DeclareRobustCommand{\VAN}[3]{#2}
\let\VANthebibliography\thebibliography
\def\thebibliography{\DeclareRobustCommand{\VAN}[3]{##3}\VANthebibliography}

\usepackage{graphicx}	
\usepackage{amsmath}	

\title[Failed SN simulations beyond BH]{Failed supernova simulations beyond black hole formation}

\author[T. Kuroda and M. Shibata]{
Takami Kuroda$^{1}$\thanks{E-mail: takami.kuroda@aei.mpg.de}
and Masaru Shibata$^{1,2}$
\\
$^{1}$Max-Planck-Institut f{\"u}r Gravitationsphysik, Am M{\"u}hlenberg 1, D-14476 Potsdam-Golm, Germany\\
$^{2}$Center for Gravitational Physics and Quantum-Information,
Yukawa Institute for Theoretical Physics, Kyoto University, Kyoto, 606-8502, Japan
}

\date{Accepted XXX. Received YYY; in original form ZZZ}

\pubyear{2015}

\begin{document}
\label{firstpage}
\pagerange{\pageref{firstpage}--\pageref{lastpage}}
\maketitle

\begin{abstract}
We present an axisymmetric failed supernova simulation beyond black hole formation, for the first time with numerical relativity and two-moment multi energy neutrino transport.
To ensure stable numerical evolution, we use an excision method for neutrino radiation-hydrodynamics within the inner part of black hole domain.
We demonstrate that our excision method is capable to stably evolve the radiation-hydrodynamics in dynamical black hole spacetime. 
As a remarkable signature of the final moment of PNS, we find the emergence of high energy neutrinos.
Those high energy neutrinos are associated with the proto-neutron star shock surface being swallowed by the central black hole and could be a possible observable of failed supernovae.
\end{abstract}




\section{Introduction}
Massive stellar collapse is one of the main formation channels of stellar-mass black hole (BH), whose existence was observationally substantiated through numerous coalescence events \citep[e.g.][]{abbott16,Abbott19}.
Massive stars heavier than $\sim8\,{\rm M}_\odot$ undergo a catastrophic gravitational core-collapse (CC) at the end stage of their evolution.
The subsequent evolutionary path is rich in variety and determines the remnant property.
Broadly speaking, less to moderately massive stars explode as core-collapse supernova (CCSN), whereas more massive stars are prone to fail the explosion, sometimes completely and sometimes exhibiting only a feeble explosion \citep{Nomoto06,Tanaka09}.
At the same time some of more massive stars are known to be accompanied by a very energetic explosion termed as hypernova \citep{iwamoto98}, whose explosion energy is about one order of magnitude larger than those of canonical SNe.

The CCSN explosion scenario and the mass range determining the fate are yet to be fully understood \citep[for reviews, see][]{Janka16,BMullerReview16,Burrows21_Review}.
It is evident, however, that unless the explosion possesses sufficient energy to expel substantial amounts of stellar mantle, the central compact remnant will ultimately acquire a mass that surpasses the maximum mass limit, above which its internal pressure cannot counteract its own self-gravitational force, thereby leading to the formation of a black hole.
The remnant property is tightly connected with its progenitor mass \citep{WHW02,Heger03}.
In general, the more massive the progenitor is, the higher the probability of being BH is.
Moreover, recent parametric studies, focusing on the explodability by the standard neutrino heating mechanism, have revealed that the compactness \citep{O'Connor11} could potentially be a good indicator of BH formation \citep[see also, e.g.,][]{Ugliano12,Sukhbold16,BMuller16,Ertl16,Ebinger19}.
Like these, the formation of a BH is predominantly determined by the compactness of the progenitor star, along with the detailed explosion scenario (but see \cite{Burrows21_Review} for counterexamples).

There are currently numerous multi-dimensional simulations reporting a successful SN explosion \citep[e.g.,][]{BMuller20,Burrows20,Stockinger20,Bollig21,Nakamura22,Vartanyan22}.
These studies are primarily directed towards less massive, or more precisely less compact, progenitor stars, in which the canonical neutrino heating mechanism can trigger the explosion, leaving behind a neutron star (NS).
However, there are several observational evidences of a ``failed'' supernova \citep{Kochanek08,Smartt15,Adams17}.
These events report a sudden disappearance of red supergiant, inferring that the whole progenitor star collapses and becomes a BH without noticeable explosions.
Furthermore exceptionally low energy SNe, e.g., SN\,2008ha \citep{Valenti09,Foley09}, were detected, which could possibly be explained by ``fallback'' during SN explosion \citep{Kawabata10,Fryer09}.
Should these events be a gravitational collapse of massive star, the remnant becomes most likely a BH due to their inferred small ejecta mass.

These observations associated possibly with a BH formation strongly motivate us to explore the failed and fallback SN scenarios.
There were, however, severe numerical difficulties in performing SN simulations in BH spacetime.
First, multi-dimensional SN simulations in general relativity (GR), for instance with numerical relativity, are still minor, e.g., \cite{BMuller10} (and its subsequent works) using the so-called conformal flatness condition (CFC) or \cite{KurodaT16} with a Baumgarte-Shapiro-Shibata-Nakamura (BSSN) formalism \citep{Shibata95,Baumgarte99}.
Since BHs are fundamentally general relativistic objects, the formation process, namely from the onset of gravitational collapse of massive progenitor to BH formation and beyond, can be precisely followed only by numerical relativity.
Second, sophisticated neutrino transport is essential for modern SN simulations.
However, numerical relativity simulation in BH spacetime combined with sophisticated neutrino transport is currently still challenging.
To date, simulations only up to BH formation \citep{KurodaT18,Shibagaki20,KurodaT22} or switching to Newtonian gravity with a large excision region (several times of the Schwarzschild radius) immediately after BH formation \citep{Chan&Muller18,Rahman22} are reported.
Very recently \cite{Sykes&BMuller23} reported the first SN simulations solving the full spatial domain above the BH, i.e., without discarding too large computational domain in the vicinity of central BH, based on the CFC metric.

The main obstacle of neutrino transport in BH spacetime, or rather immediately after BH formation, stems from the rapid change of matter field.
At the moment of BH formation, the (rest mass) density just above the BH is generally high $\gtrsim10^{14}$\,g\,cm$^{-3}$.
The density, however, quickly decreases to $\sim10^{10}$\,g\,cm$^{-3}$ within a few ms concomitantly with the proto-neutron star (PNS) being swallowed by the central BH.
This indicates that the region in the vicinity of the BH rapidly shifts from optically thick to thin condition and such extreme condition makes neutrino transport with full interactions a significantly challenging subject.
In addition, the matter (and probably also radiation) field inside the BH is typically required to be ``excised'' for stable numerical evolution.
As of now, however, there is no concrete method how we should treat the radiation field inside the excised region and also inside BH for stable numerical evolution.

In this study, we report our first SN simulation beyond BH formation with numerical relativity and multi-energy neutrino transport.
We use an excision method for both matter and neutrino radiation fields inside a part of BH domain.
Our excision method demonstrates stable evolution immediately after BH formation as well as in the subsequent BH accretion phase.
Furthermore, we find the emergence of high energy neutrinos associated with the PNS shock surface being swallowed by the central BH, which could potentially be a probe of the very final moment of PNS.
We also show that these high energy neutrinos could be detectable by the current and next-generation neutrino detectors if the BH formation happens in our Galaxy.

This paper is organized as follows.
Section~\ref{sec:Method} starts with a concise summary of our GR radiation-hydrodynamic scheme with an excision scheme and also describe the initial setup of the simulation.
The main results and detailed analysis of our new findings are presented in Section~\ref{sec:Results}.
We summarize our results and conclude in Section~\ref{sec:Summary}.
Throughout the paper, Greek indices run from 0 to 3 and Latin indices from 1 to 3, except $\nu$ and $\varepsilon$ which denote neutrino species and energy, respectively.

\section{Method}
\label{sec:Method}

In our full GR radiation-hydrodynamics simulations, we solve the evolution equations of metric, hydrodynamics, and energy-dependent neutrino radiation.
Each of the evolution equations is solved in an operator-splitting manner, while the system evolves selfconsistently as a whole, satisfying the Hamiltonian and momentum constraints \citep{KurodaT16}.
In Sec.~\ref{sec:Radiation hydrodynamics in BH spacetime}, we describe our numerical method focusing particularly on the excision method applied to the neutrino radiation-hydrodynamics variables.
Sec.~\ref{sec:Model} is devoted to explaining the computed model and numerical setup.

\subsection{Radiation hydrodynamics in BH spacetime}
\label{sec:Radiation hydrodynamics in BH spacetime}
We solve full GR multi-energy neutrino transport equations in axisymmetric $2+1$ dimensions (two spatial dimensions and one momentum-space dimension).
Details of the code are described in our previous studies \citep{KurodaT16,KurodaT22}.
The black hole spacetime is evolved using the BSSN formalism \citep{Shibata95,Baumgarte99} with a fourth order finite differencing for the spatial derivatives and a four-step Runge-Kutta method.
We choose `1+log' slicing condition for the lapse and gamma-driver condition for the shift vector \citep{Alcubierre03}.
BH formation is determined by identifying the location of apparent horizon (AH) by an AH finder, e.g., \cite{ShibataAH97}.
After the AH formation, we enforce an excision method for radiation-hydrodynamics inside the AH, while we evolve the full black hole spacetime without excision for geometrical variables.

Here we will briefly explain our excision technique for radiation-hydrodynamics.
Once the AH is found, we divide the interior of AH into two: inner and outer regions.
The interface of these two regions is locating at $fr_{\rm AH}(\theta)$, where $f\in[0,1]$ and $r_{\rm AH}(\theta)$ denotes the radius of AH at $\theta$-direction with $\theta$ being the angle with respect to $z$-axis.
In the outer region, we solve the full neutrino radiation-hydrodynamics in the same way as the outside of AH (i.e. $r>r_{\rm AH}$).
On the other hand, we excise the inner region and artificially set all primitive variables, i.e., the rest mass density $\rho$, entropy $s$, electron fracion $Y_e$, spacial components of four-velocity $u^i$, and the zeroth and first order neutrino radiation moments $(E_{(\nu,\varepsilon)},{F^i}_{(\nu,\varepsilon)})$, as 
\begin{eqnarray}
\label{eq:primitiveV}
\left[
\begin{array}{c}
\rho \\
u^i \\
s \\
Y_e \\
E_{(\nu,\varepsilon)} \\
{F_{(\nu,\varepsilon)}}_i \\
\end{array}
\right]=
\left[
\begin{array}{c}
\sim0.1\rho_{\rm max} \\
0 \\
\approx 1.5\,k_{\rm B}\,{\rm baryon^{-1}} \\
\approx 0.15 \\
{E_{\rm thick}}_{(\nu,\varepsilon)} \\
{{F_{\rm thick}}_{(\nu,\varepsilon)}}_i \\
\end{array}
\right]\,\,{\rm for}\,\,r(\theta)\le fr_{\rm AH}(\theta).
\end{eqnarray}
Here $\rho_{\rm max}$ represents the maximum rest mass density outside of the AH, which therefore changes its value with time due to the mass accretion onto BH.
Regarding the entropy and electron fraction, we use fixed values taken from typical NS structures.
The zeroth and first order radiation moments $({E_{\rm thick}}_{(\nu,\varepsilon)},{{F_{\rm thick}}_{(\nu,\varepsilon)}}_i)$ inside the inner region are enforced to be the moments in the optically thick limit \citep[c.f. Eqs.~(6.14)--(6.15) in][]{Shibata11} assuming the beta  equilibrium with matter.

We shortly touch the appropriate value for $f$.
Usually, source terms for neutrino-matter interactions including gravitational red-shift and Doppler terms are quite {\it stiff}.
Inside the inner region $r(\theta)\le fr_{\rm AH}(\theta)$, we do not evolve any radiation-matter fields, that is, these stiff source terms are suddenly switched off across the excision boundary.
Such artificial treatment inevitably causes spurious behaviours appearing especially in the radiation fields near the excision boundary.
If we choose the value of $f$ to be close to unity, those spurious oscillations eventually propagate even out to the outside of AH and the simulation will be crashed.
Therefore in this study we set $f=0.5$ to avoid such pathological behavior.
With these treatments, we found numerically stable neutrino radiation-hydrodynamic evolution in BH spacetime.

\subsection{Model}
\label{sec:Model}
We use a non-rotating massive star with zero metallicity, whose initial mass at its zero-age main sequence is $70\,{\rm M}_\odot$ \citep{Takahashi14}.
It has a substantially high compactness parameter $\xi_{2.5}=1$ \citep{O'Connor11} at the final evolution phase.
This progenitor star was reported to form a BH within a few hundred milliseconds after the first bounce \citep{KurodaT18,Shibagaki21}.
We use the DD2 EOS of \cite{Typel10}.
The maximum NS mass of DD2 for cold and non-rotating case is 2.42\,M$_\odot$, which is consistent with the existence of observationally confirmed massive NSs with $\sim2$\,M$_\odot$ \citep{Demorest10,Antoniadis13,Cromartie20}.

The 2D axially symmetric computational domain extends to $1.5\times10^4$\,km from the center.
In the cylindrical computational domain, 2:1 ratio nested boxes with 11 refinement levels are embedded, and each nested box contains $64\times 64$ cells so that the finest resolution at the center becomes $\approx$230\,m. In this work, we assume the plane symmetry with respect to the equatorial plane. The neutrino energy space $\varepsilon$ logarithmically covers from 3 to 400\,MeV with 14 energy bins.
In this study, we use the up-to-date neutrino rates of \citet{Kotake18}, which are used also in our recent studies \citep{KurodaT22,KurodaT23}.

\section{Results}
\label{sec:Results}
\begin{figure}
\begin{center}
\includegraphics[angle=-90.,width=\columnwidth]{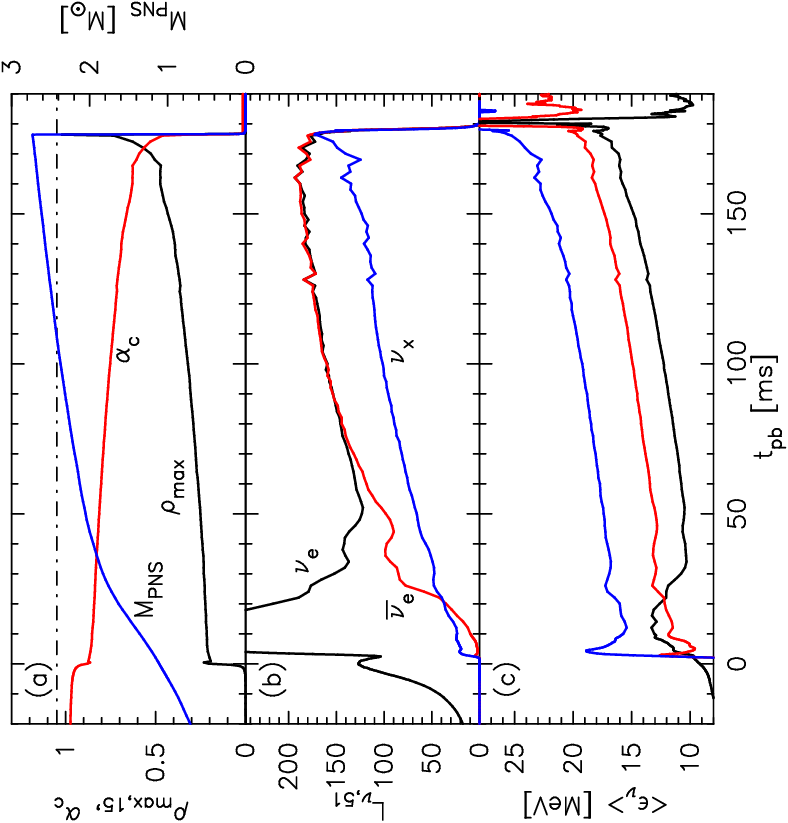} 
\caption{Overall evolution feature.
Panel (a): the maximum rest-mass density $\rho_{\rm max}$ (black), central lapse function $\alpha_{\rm c}$ (red), and baryon mass of PNS $M_{\rm PNS}$ (blue); (b): neutrino luminosity $L_{\nu,51}$ in units of $10^{51}$\,erg\,s$^{-1}$; and (c) neutrino mean energy $\langle\varepsilon_\nu\rangle$.
Neutrino profiles are evaluated at $r=400$\,km.
In panels (b) and (c), the color represents neutrino species: electron type neutrino (black), electron type antineutrino (red), and heavy lepton type neutrino (blue).
\label{fig:Overall}}
\end{center}
\end{figure}

We first describe the picture of post-bounce evolution till the formation of BH.
Fig.~\ref{fig:Overall} shows: (a) the maximum rest-mass density $\rho_{\rm max,15}$ in units of $10^{15}$\,g\,cm$^{-3}$ (black), baryon mass of PNS $M_{\rm PNS}$ (blue), and central lapse function $\alpha_{\rm c}$ (red); (b) neutrino luminosity $L_{\nu,51}$ in units of $10^{51}$\,erg\,s$^{-1}$ for neutrino species; and (c) neutrino mean energy $\langle\varepsilon_\nu\rangle$.
The PNS surface is defined by the location for which the rest mass density drops below $10^{10}$\,g\,cm$^{-3}$.
$L_\nu$ and $\langle\varepsilon_\nu\rangle$ are evaluated from the emergent neutrino spectra measured at $r=400$\,km.
In panel (a), we also plot the maximum mass of DD2 EOS for cold and non-rotating stars by the horizontal dash-dotted line of 2.42\,M$_\odot$.

Panel (a) exhibits that the $M_{\rm PNS}$ exceeds the maximum allowed mass of current EOS at $t_{\rm pb}\sim100$\,ms.
However, because of an additional contribution from thermal pressure, the PNS does not immediately collapse to a black hole.
From the maximum density evolution, we see a sharp increase at $t_{\rm pb}\sim177$\,ms, at the same time $\alpha_c$ decreases to $\sim0$. This signals the BH formation.
Prior to the BH formation at $t_{\rm pb}\gtrsim 160$\,ms, electron and anti-electron type neutrino luminosities show a decresing trend, while heavy-lepton neutrinos show a rapid increase in both its luminosity and mean energy.
These features were previously identified in 1D full-GR simulations with Boltzmann neutrino transport \cite{Liebendorfer04} and are commonly observed in the literature, due to rapid contraction of the PNS to the forming BH (see also, \cite{Sumiyoshi07,Fischer09,Hempel12,Gullin&O'Connor22} as well as 3D models by \cite{KurodaT18,Shibagaki21}).
The overall features before the BH formation are in a good agreement with our former model $z70$ reported in \cite{KurodaT22}, in which the DD2-based nuclear EOS taking into account a first-order quantum chromodynamics (QCD) phase transition was used.
Taking into account the fact that the QCD phase transition occurs after the PNS starts collapsing~\citep{KurodaT22}, the agreement between the current and previous models is quite reasonable.

We also compare BH formation time with previous related studies.
\cite{O'Connor11} presented a nice correlation between BH formation time, obtained from various 1D GR models, and compactness parameter of progenitor star.
According to their Fig.~6, massive stars having $\xi_{2.5}=1$, which is the case for the current model, are forming BH at $t_{\rm pb}\sim250-750$\,ms, where the time variation reflects the different nuclear EOS.
\cite{Powell21} performed faint SN simulations in 3D using a zero-metallicity progenitor star with $85$\,M$_\odot$, whose compactness parameter is $\xi_{2.5}=0.86$.
They witnessed shock revival prior to BH formation, which to some extent suppresses subsequent mass accretions onto the PNS and may delay the BH formation.
Their models exhibited BH formation occurring at $t_{\rm pb}\sim290-590$\,ms.
Using similar massive progenitor stars, \cite{Rahman22} also demonstrated faint SN scenarios with BH formation occuring at $t_{\rm pb}\sim350-400$\,ms.
In addition, a recent study of \cite{Sykes&BMuller23} presented BH formation at $t_{\rm pb}\sim220$\,ms for the same progenitor model used in \cite{Powell21}.
Considering that our numerical formalism is totally independent from these previous studies and also that we use a different progenitor model, some time variations in BH formation time are expected to emerge.
At the same time, comparing to less massive stars, e.g., with $\xi_{2.5}\sim0.25$, which are predicted to form BH at $t_{\rm pb}\gtrsim$2\,s \citep{O'Connor11}, unless successful shock revival does not occur, all previous studies including this study are presenting consistent BH formation time, i.e., substantially quicker than $t_{\rm pb}\gtrsim$2\,s expected in less massive stars.

\begin{figure}
\begin{center}
\includegraphics[angle=-90.,width=\columnwidth]{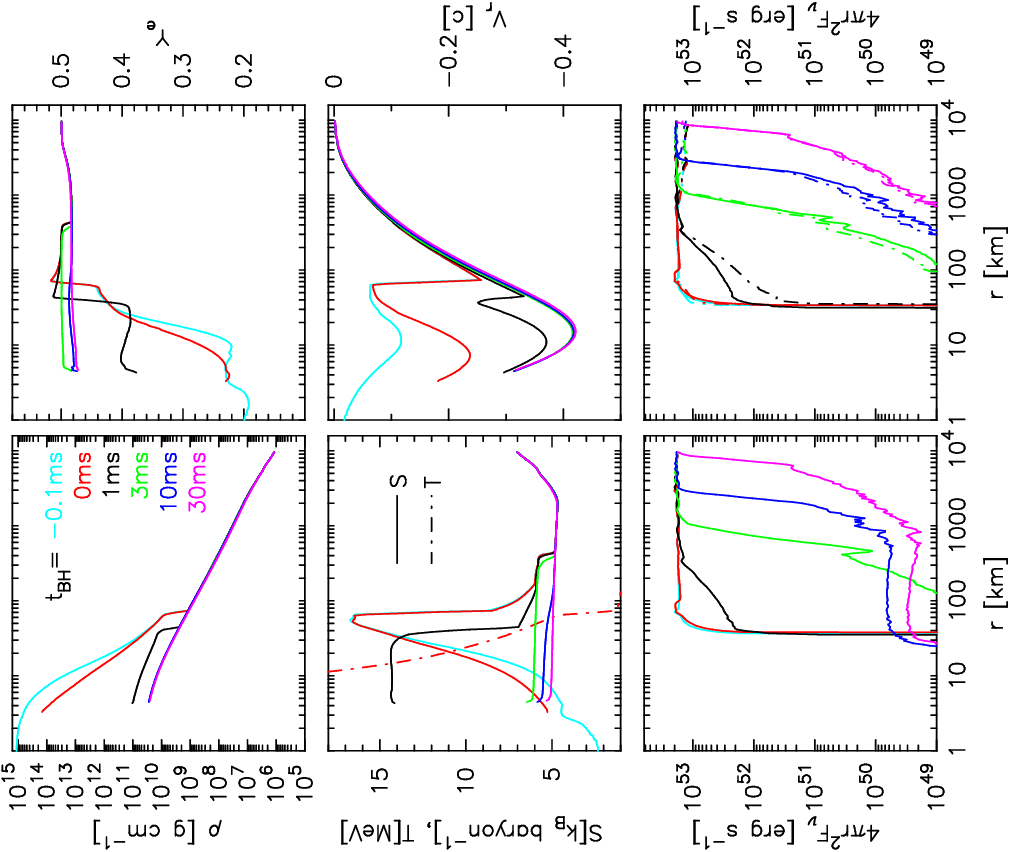} 
\caption{Spherically averaged radial profiles of the rest mass density $\rho$ (top-left), electron fraction $Y_e$ (top-right), entropy per baryon $s$ (middle-left), radial component of the three-velocity $v^r\equiv u^r/u^t$ (middle-right), neutrino luminosity for $\nu_e$ (bottom-left), $\bar\nu_e$ (solid, bottom-right), and $\nu_x$ (solid-dashed, bottom-right) at different times denoted in the top-left panel.
In the middle-left panel, we also plot a temperature profile, but only at $t_{\rm BH}=0$\,ms (red dash-dotted line).
\label{fig:SphProf}}
\end{center}
\end{figure}

\begin{figure}
\begin{center}
\includegraphics[angle=-90.,width=\columnwidth]{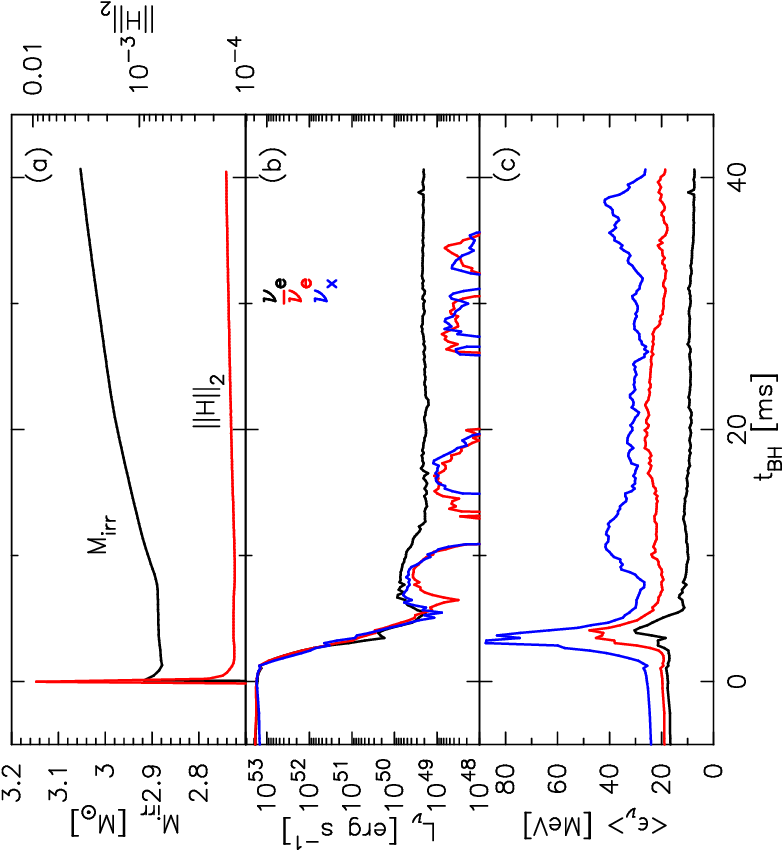} 
\caption{Post BH formation evolution of: (a) the irreducible mass $M_{\rm irr}$ and 2-norm of Hamiltonian constraint violation $||H||_2$, (b) neutrino luminosities, and (c) mean neutrino energies, as a function of $t_{\rm BH}$.
In panels (b) and (c), the color represents neutrino species: electron type neutrino (black), electron type antineutrino (red), and heavy lepton type neutrino (blue).
\label{fig:Overall_BH}}
\end{center}
\end{figure}

Next we discuss the neutrino radiation-hydrodynamics evolution after the BH formation, focusing mainly on how effectively our excision method manage to prevent propagation of spurious behaviours often appeared at the excision boundary.
Fig.~\ref{fig:SphProf} displays spherically averaged spatial profiles of the rest mass density (top-left), electron fraction (top-right), entropy (middle-left), radial component of the three velocity (middle-right), electron type neutrino luminosity (bottom-left), and anti-electron type (solid-line) and heavy-lepton type (dash-dotted line) neutrino luminosities (bottom-right), at several time slices.
In the middle-left panel, we supplementary plot a temperature profile, but only at the formation of BH (red dash-dotted line), which is used in the later discussion with Fig.~\ref{fig:Overall_BH}.
Each color represents the post BH formation time $t_{\rm BH}$, denoted in the top-left panel. Once the AH is formed, we plot structures only aoutside the AH.

Slightly before AH formation at $t_{\rm BH}=-0.1$\,ms, the central density exceeds $10^{15}$\,g\,cm$^{-3}$ and the velocity profile inside the PNS shows the infalling structure.
For $t_{\rm BH}\ge0$\,ms, for which we apply an excision method described in the previous section, we see essentially no numerical instabilities at the interface of the AH.
All the neutrino radiation fields and hydrodynamical variables exhibit smooth structures across the AH and subsequently swallowed into its inside.
From the density structural evolution, the maximum density drops by four orders of magnitude, from $\sim10^{14}$\,g\,cm$^{-3}$ to $\sim10^{10}$\,g\,cm$^{-3}$, within a few ms, presenting a clear transition from optically thick to thin conditions.
This feature makes SN simulations in dynamical BH spacetime one of numerically challenging subjects.
We found that, if we suddenly switch off the neutrino-matter interactions inside the AH, it causes spurious behaviors, which eventually leak out to the outside and lead to a code crash.
Therefore we believe that it is essential to ensure a buffer zone between the AH and the excised region, especially when the neutrino radiation fields are taken into account.
During the first few ms after AH formation, low-$Y_e$ and high entropy material, which represent typical PNS shocked material, are still present outside the AH.
They are, however, immediately swallowed by the BH and for $t_{\rm BH}\gtrsim3$\,ms the BH accretion enters a nearly steady state, exhibiting high-$Y_e$ ($\sim0.49$) and relatively low entropy ($\sim5$\,k$_{\rm B}$\,baryon$^{-1}$) flows (see magenta lines).

Next we focus on how the neutrino signals in association with the BH formation are radiated away.
Bottom two panels indicate that all neutrino species have an outgoing flux for  $r\gtrsim30$\,km at the time of the BH formation.
In the vicinity of AH, on the other hand, neutrino radiation fields experience a strong drag by infalling high density component ($\gtrsim10^{12}$\,g\,cm$^{-3}$) and have an inward flux.
After the mass accretion becomes a nearly steady state flow for $t_{\rm BH}\gtrsim3$\,ms, the dominant neutrino-matter interaction is the electron capture due to continuous replenishment of high-$Y_e$ materials ($\sim0.49$, see top-right panel) from stellar mantle.
It results in a sustained neutrino emission even after the BH formation for electron type neutrinos (see blue and magenta lines in the bottom-left panel in Fig.~\ref{fig:SphProf}), while the rest of neutrino species has essentially no production channel and their neutrino luminosities quickly subside.
\cite{Sykes&BMuller23} reported a BH excision scheme with neutrino transport.
According to their long time failed CCSN simulation in 1D spherical symmetry, qualitatively similar spatial profiles of neutrino luminosities, namely relatively strong $\nu_e$ emission continuing even after BH formation, was also reported.

Fig.~\ref{fig:Overall_BH} displays: (a) the irreducible mass $M_{\rm irr}$ and 2-norm of Hamiltonian constraint violation $||H||_2$, (b) neutrino luminosities, and (c) mean neutrino energies, as a function of $t_{\rm BH}$.
Here, $M_{\rm irr}$ is defined by the area of apparent horizon $A$ as $M_{\rm irr}=\sqrt{A/16\pi}$ \citep[cf.][]{Baumgarte_AH,ShibataAH97} and $||H||_2$ measures the constraint violation only for numerical cells outside the AH.
From panel (a), the irreducible mass shows an increasing trend from $M_{\rm irr}\sim2.88$\,M$_\odot$ to $\sim3.06$\,M$_\odot$ during the first 40\,ms.
At the moment of the AH formation, the measured value of the protoneutron star mass, $M_{\rm PNS}$, is $\sim2.76$\,M$_\odot$, which rapidly decreases to $\lesssim0.001$\,M$_\odot$ (the total mass outside of the AH and where $\rho\ge10^{10}$\,g\,cm$^{-3}$ is met) within a few ms.
It means that the estimated $M_{\rm irr}$ is slightly larger than $M_{\rm PNS}$ at $t_{\rm BH}=0$\,ms.
Furthermore, from panel (a), $M_{\rm irr}$ initially shows a slightly odd behavior, a nearly constant evolution until $t_{\rm BH}\sim8$\,ms, and it increases afterward.
From these, we naively suspect that the current numerical resolution at the center $\Delta x\sim230$\,m might not be high enough\footnote{The BH is resolved by $\sim13-14$ grid points at its formation.} to accurately resolve the location of apparent horizon and may tend to overestimate the initial BH mass approximately by $\sim0.1$\,M$_\odot$, i.e., $\sim3$\,\% error in the evaluation for the total BH mass or the AH radius.
However, once the system relaxes to a quasi-steady state for $t_{\rm BH}\gtrsim10$\,ms, $M_{\rm irr}$ increases with a reasonable growth rate of $\dot M_{\rm irr}\approx 4.66$\,M$_\odot$\,s$^{-1}$, which agrees approximately with that of the PNS mass, $\dot M_{\rm PNS}\approx 4.73$\,M$_\odot$\,s$^{-1}$, before the BH formation (see panel (a) in Fig.~\ref{fig:Overall}).
The 2-norm of Hamiltonian constraint $||H||_2$ stays around $\sim10^{-4}$ without any secular increase after BH formation.

Regarding the neutrino signals, the neutrino luminosity for all species show a rapid distinction and eventually migrate to a quasi steady state for $t_{\rm BH}\gtrsim5$\,ms.
From panel (b), $L_{\nu_e}$ stays around $\sim2\times10^{49}$\,erg\,s$^{-1}$ till the end of our calculation, which features a long term steady state mass accretion onto the BH.
Nearly constant $L_{\nu_e}$ of the order of $\mathcal O(10^{49})$\,erg\,s$^{-1}$ is also reported in \cite{Sykes&BMuller23}.

The neutrino mean energy $\langle \varepsilon_\nu \rangle$ may reveal the final moment of devastating PNS collapse.
As can be clearly seen, $\langle \varepsilon_\nu \rangle$ for all neutrino species show a drastic increase at $t_{\rm BH}\sim3$\,ms.
This is particularly the case for heavy lepton type neutrinos, which show a remarkably high mean energy of $\langle \varepsilon_{\nu_x} \rangle\sim90$\,MeV.
These values are even higher than those from the QCD CCSN models \citep{Fischer18,KurodaT22}, which are also known to emit high energy neutrinos $\langle \varepsilon_{\nu_x} \rangle\sim40$\,MeV due to strong shock heating in association with the quark core bounce.
We will now shortly discuss their possible excitation mechanism.
First, since we measure the emergent neutrino signals at $r=400$\,km, these high energy neutrinos are produced at $t_{\rm BH}\sim1-2$\,ms.
From Fig.~\ref{fig:SphProf}, this time corresponds exactly to the time when huge amounts of hot PNS envelope together with a shock surface infall with a relativistic speed of $\sim0.3c$.
The highest temperature of collapsing PNS material (middle-left panel in Fig.~\ref{fig:SphProf}) for the regions of $r\gtrsim30$\,km, where $F_{\nu_x}$ has a positive sign (bottom-right panel) and can contribute to the emergent neutrino spectrum, is merely $T\sim10$\,MeV.
It indicates that heavy lepton type neutrinos, whose energy are $\langle \varepsilon_{\nu_x} \rangle\sim30$\,MeV, could be barely explained via such as pair production channel, although it is not likely for much higher neutrino energy of $\sim90$\,MeV.

\begin{figure*}
\begin{center}
\includegraphics[angle=-90.,width=2.0\columnwidth]{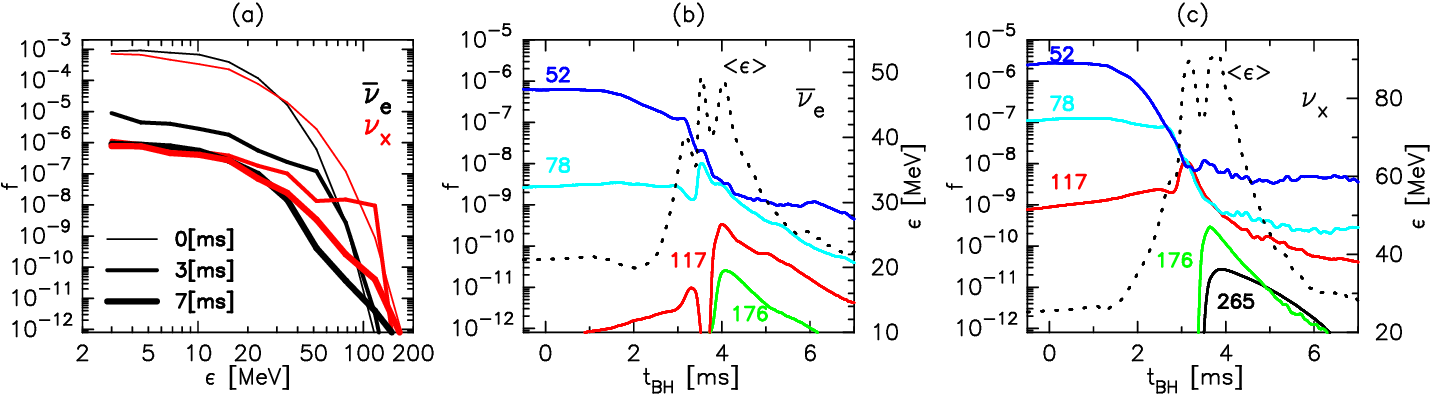} 
\caption{From left: (a) the distribution function $f_\varepsilon$ for $\bar\nu_e$ (black lines) and $\nu_x$ (red lines) at three different time slices: $t_{\rm BH}=0$\,ms, 3\,ms (corresponding to the time when high energy neutrinos are observed), and 7\,ms, (b) time evolution of distribution function $f_{\varepsilon}$ for all energy bins higher than $\varepsilon\ge52$\,MeV (this time, 52, 78, 117, 176, and 265\,MeV) (solid lines) and mean energy $\langle \varepsilon\rangle$ (dashed line) for $\bar\nu_e$, and (c) same as the panel (b) but for $\nu_x$.
All these values are measured at $r=400$\,km.
\label{fig:NuSpec}}
\end{center}
\end{figure*}
To further discuss their origin, we examine their spectral features. 
Fig.~\ref{fig:NuSpec} depicts: (a) the distribution function $f_\varepsilon$\footnote{We reconstruct the distribution function $f_\varepsilon$ simply by $f_\varepsilon=J_\varepsilon/4\pi\varepsilon^3$, where $J_{\varepsilon}$ denotes the zeroth order neutrino radiation moment measured in the comoving frame at the energy bin $\varepsilon$. With an appropriate closure relation, $J_\varepsilon$ is determined from the zeroth and first order radiation momenta $(E_\varepsilon,F_\varepsilon^\mu)$, which are measured in the Eulerian frame and are the basic variables evolved in our M1 neutrino transport.} for $\bar\nu_e$ (black lines) and $\nu_x$ (red lines) at three different time slices: $t_{\rm BH}=0$\,ms, 3\,ms (corresponding to the time when high energy neutrinos are observed), and 7\,ms, (b) time evolution of distribution function $f_{\varepsilon}$ for all energy bins higher than $\varepsilon\ge52$\,MeV (this time, 52, 78, 117, 176, and 265\,MeV) (solid lines) and mean energy $\langle \varepsilon\rangle$ (dashed line) for $\bar\nu_e$, and (c) same as the panel (b) but for $\nu_x$.
All these values are measured at $r=400$\,km.

From panel (a), the energy spectrum at $t_{\rm BH}=3$\,ms for $\nu_x$ exhibits a flatter profile with relatively more populations for neutrinos with $\gtrsim50$\,MeV.
Such feature cannot be seen in other two time snapshots.
We attribute the flatter profile to a consequence of more effective isoenergy scatterings taking place in the upstream to the relativistically infalling shock surface.
Because of the rapid infall of the PNS shock surface (see $v_r$-profiles from $t_{\rm BH}=-0.1$\,ms to 1\,ms in Fig.~\ref{fig:SphProf}), the outgoing comoving neutrino flux ahead of the shock becomes relatively larger.
Consequently the effect of isoenergy neutrino scatterings becomes more prominent compared to the case with a stationary shock surface.
Furthermore, that impact is more visible for high energy neutrinos as the cross section of the isoenergy scatterings is proportional to the square of the incoming neutrino energy.
Indeed, from panel (c), the distribution function for heavy lepton type neutrinos shows an increase(decreasing) trend for $\varepsilon\ge117(\le78)$\,MeV at $t_{\rm BH}\lesssim3$\,ms.
Particularly at the energy bin $\varepsilon=117$\,MeV ($f_{\varepsilon=117}$: red line), its increase is noteworthy with its maximum appearing at $t_{\rm BH}\sim3$\,ms.
Neutrinos at higher energie bins ($\varepsilon=176$ and 265\,MeV) also show a sudden increase with slight time delays of $\sim0.5$\,ms from the peak time for $f_{\varepsilon=117}$.
These time delays are mostly due to that higher energy neutrinos require a longer time for escaping from collapsing stellar mantle.
On the other hand, regarding $\bar\nu_e$ (as well as $\nu_e$), the less population of high energy neutrinos ($\varepsilon\gtrsim50$\,MeV) prior to the BH formation than that of $\nu_x$ (compare two thin lines in panel (a)) leads simply to a less noticeable increase at $t_{\rm BH}\sim3-4$\,ms.
Additionally, the presence of charged current reactions tend to suppress their increase.
In fact, $f_{\varepsilon\ge117}$ for $\bar\nu_e$ shows approximately an order of magnitude smaller values than that for $\nu_x$.
These features result in the observed high energy neutrinos pronounced for heavy lepton type ones (Fig.\ref{fig:Overall_BH}).
Although our moment formalism cannot capture the particle acceleration mechanisms at the shock front, non-thermal shock acceleration \citep{Kazanas81,Giovanoni89,Nagakura&Hotokezaka21} is also reported to excite high energy neutrinos from CCSNe.

As a comparison with previous studies, \cite{Gullin&O'Connor22} has perofrmed a GR Monte Carlo neutrino transport and reported high energy neutrinos with $\langle \varepsilon_{\nu_x} \rangle\sim50$\,MeV in association with BH formation.
Since their calculations are performed on the fixed spacetime and matter fields after BH formation, quantitative differences in $\langle \varepsilon_{\nu}\rangle$ from ours are inevitable.
We, however, believe that the emission of high energy neutrinos just after the BH formation seem to be a common feature and might be used as a smoking gun of infall of PNS surface.
\cite{Rahman22} performed CCSN simulations with BH formation.
However, since they excise the innermost $400$\,km once they find the AH and also their models present a successful shock expansion, i.e., corresponding to the fallback SN model, the emergence of high energy neutrinos similar to ours was not reported.

\begin{figure}
\begin{center}
\includegraphics[angle=-90.,width=\columnwidth]{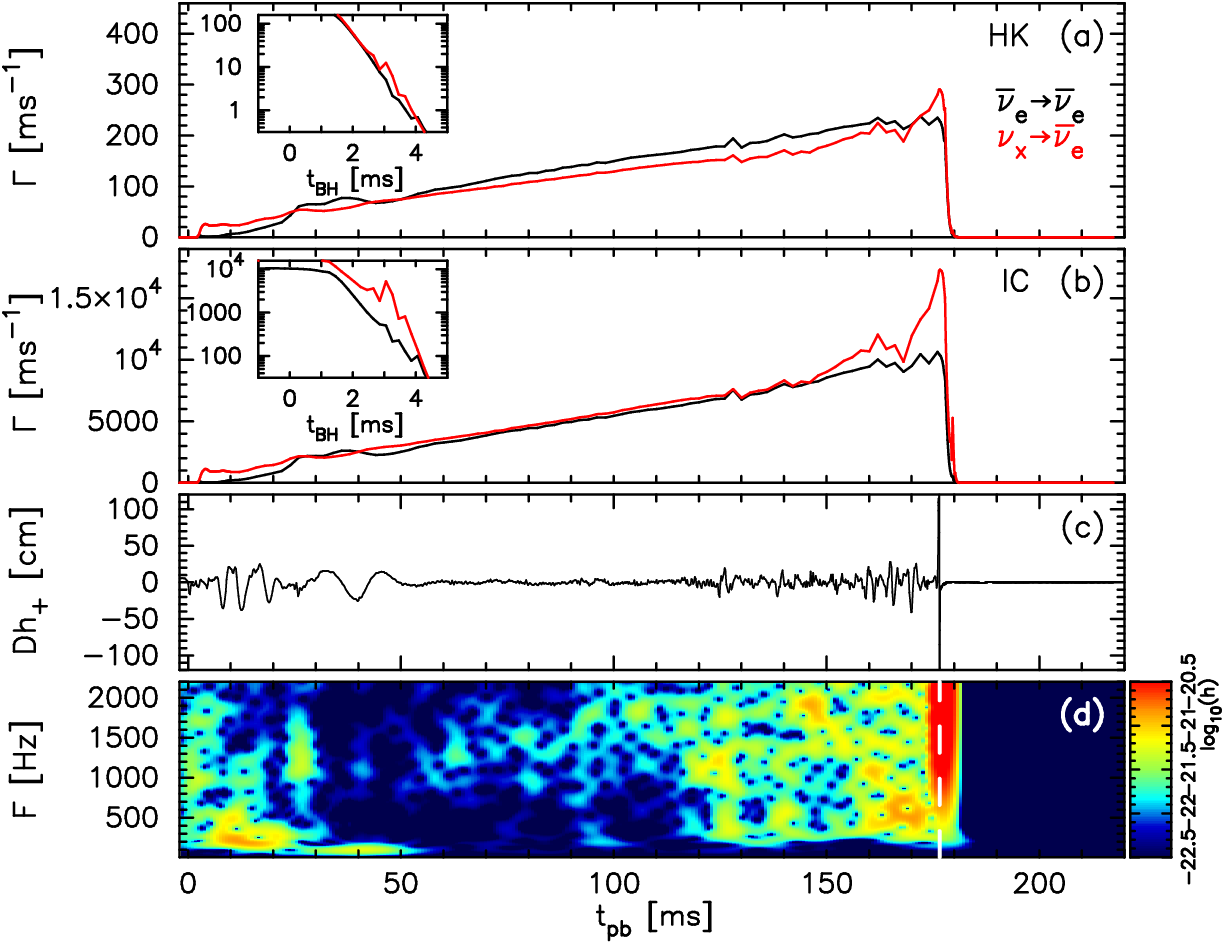} 
\caption{From top: (a) the neutrino detection rate $\Gamma$ of Hyper-Kamiokande (HK); (b) $\Gamma$ of IceCube (IC); (c) matter origin GWs $Dh_+$; and (d) spectrogram of $h_+$ obtained by a short-time Fourier transform.
We assume a source distance of $D=10$\,kpc.
\label{fig:GW_NC}}
\end{center}
\end{figure}
Finally, we discuss observable multi messenger signals for a current failed CCSN model.
Fig.~\ref{fig:GW_NC} displays from top: (a) the neutrino detection rate $\Gamma$ of Hyper-Kamiokande (HK) \citep{abe11,HK18}; (b) $\Gamma$ of IceCube (IC) \citep{abbasi11,salathe12}; (c) matter origin gravitational waves (GWs) $Dh_+$; and (d) spectrogram of $h_+$ obtained by a short-time Fourier transform.
We assume a source distance of $D=10$\,kpc.
$h_+$ is the gravitational wave strain, which is calculated from a standard quadrupole formula, and we show only the non-vanishing component in axisymmetric profile observed along the equatorial plane.
The neutrino detection rate $\Gamma$ is evaluated in the same way as \cite{KurodaT22} assuming a Fermi-Dirac distribution for the neutrino energy spectrum \citep{Lund10,Takiwaki18}.
Note that in the evaluation for $\Gamma$, we consider two extreme cases: all $\bar\nu_e$ emitted from the source reach the detectors without neutrino flavor conversion and cause the signal at the detectors (black lines in the figure); all $\bar\nu_x$ (identical to $\nu_x$ in this study) emitted from the source are completely swapped by $\bar\nu_e$ and cause the signals (red lines).
In inset of the upper two panels, we show a magnified view of $\Gamma$ relative to BH formation time $t_{\rm BH}$ to feature detection of high energy neutrinos.

Regarding the neutrino detection rate $\Gamma$, both of the two extreme cases, i.e., with and without neutrino flavor conversion, essentially show a quantitatively similar monotonic increase until the BH formation.
This feature can be seen for both detectors.
This indicates that the possible range of neutrino oscillation effects \citep[see][for a review]{Mirizzi16}, i.e. the region bounded by two lines in panels (a,b), is quite small, compared to previous studies using less massive progenitor stars \citep[e.g.][]{Tamborra12,KurodaT22}.
For instance, $\Gamma_{\bar\nu_e\rightarrow \bar\nu_e}$ becomes $\sim1.5$ times higher than $\Gamma_{\bar\nu_x\rightarrow \bar\nu_e}$ for $t_{\rm pb}\gtrsim100$\,ms for CCSN models with less massive progenitor stars, while the current one with a more massive progenitor star presents roughly comparable values.
Another remarkable feature is rapid increase of $\Gamma_{\bar\nu_x\rightarrow \bar\nu_e}$ (red lines) as the PNS approaches BH formation ($t_{\rm pb}\gtrsim150$\,ms).
It is a clear signature of the increasing behavior of both $L_{\nu_x}$ and $\langle \varepsilon_{\nu_x}\rangle$ shown in Fig.~\ref{fig:Overall}.
We also discuss if the high energy heavy lepton type neutrinos, as a possible signature of the shock surface being swallowed by BH, could be observed.
From insets, we can marginally observe a slight increase for $\Gamma_{\bar\nu_x\rightarrow \bar\nu_e}$ (red lines) at $t_{\rm BH}\sim3$\,ms, which is more visible for IC.
This time is consistent with the emission time of high energy neutrinos (see panel (c) in Fig.~\ref{fig:Overall_BH}).
If we could observe such a tentative increase of neutrino detection during the exponential decay, it could be a possible signature of the aforementioned final moment of the PNS shock surface.

Bottom two panels show the emitted GWs.
We see essentially the same features as have been discussed for model $z70$ in \cite{KurodaT22}.
During the first $\sim50$\,ms after bounce, relatively large and low frequency GWs originated from postbounce convective motions are observed, whose amplitudes and frequencies reach $\sim50$\,cm and $\sim100$\,Hz, respectively.
Afterward the gravitational waveform shows a considerable subsidence, which is then disrupted at $t_{\rm pb}\gtrsim120$\,ms.
At the moment of BH formation, burst like GWs of the order of $\sim100$\,cm are emitted presenting a broad band emission.
Once the BH is formed and BH accretion settles into a quasi steady state for $t_{\rm BH}\gtrsim3$\,ms, we observe essentially no GWs for the current non-rotating model.
As a comparison to a previous 2D GR study \citep{Rahman22}, which performed faint SN simulations using an $80$\,M$_\odot$ progenitor star, the current GWs are showing consistent behaviors in the initial convection phase ($t_{\rm pb}\lesssim50$\,ms).
During this phase, the amplitude and typical frequency reach $Dh\sim30-40$\,cm and $F\sim100$\,Hz, respectively, in their non-rotating model.
These values are quite consistent with our findings.
Although a direct comparison in the subsequent phase ($t_{\rm pb}\gtrsim50$\,ms till BH formation) may not be so meaningful, as their models are faint SN, i.e., exhibiting shock revival before BH formation, high frequency GWs ($F\sim1000$\,Hz) are also observed prior to BH formation, which could potentially be another common feature.

\section{Summary}
\label{sec:Summary}

We have presented a results of 2D axisymmetric CCSN simulation for a massive star with $70$\,M$_\odot$.
Our core-collapse supernova model is based on numerical relativity, which solves the GR neutrino-radiation hydrodynamics equations together with the two-moment (M1) neutrino transport equations of \citet{KurodaT16}.
We used up-to-date neutrino opacities following \cite{Kotake18} and employed the DD2 EOS of \citet{Typel10}.
In this framework, we follow for the first time ``beyond BH formation''.
To ensure stable numerical evolution, we use an excision method for neutrino radiation-hydrodynamics, while we evolve the geometrical variables for entire computational domain.

Our results showed consistent PNS evolution and multi-messenger signals during the PNS contraction phase with previous studies, for which the same progenitor model was used \citep{KurodaT18,Shibagaki21,KurodaT22}.
The current non-rotating PNS model exceeds the maximum NS mass for DD2 EOS at $\sim100$\,ms after bounce.
Subsequently, it initiates the second gravitational collapse, resulting in BH formation at $t_{\rm pb}\sim177$\,ms.
After we identify the AH, our excision technique demonstrates its capability to stably evolve the radiation-hydrodynamics in dynamical BH spacetime.
We solve the full neutrino-matter interactions taking into account the gravitational redshift and Doppler terms from the AH down to the excision domain, so that spurious oscillations often appearing around the excision surface do not leak outside the AH.
We also mention that our current numerical method satisfies the Hamiltonian constraint well and its violation after BH formation is free from secular growth.

After the BH formation, the PNS envelope was simply swallowed by the BH and the system transitions to a nearly steady BH-accretion phase within a few ms.
Afterward the BH mass, i.e. the area of AH, gradually increases because of the continuous mass inflow.
The accretion flow is composed of high-$Y_e$ $(\sim0.5)$ material, reflecting the component of progenitor core (i.e. iron).

On the contrary to the simple collapse dynamics of PNS, its impact on the emergent neutrino signals was not so trivial.
Our findings are: (1) neutrinos with significantly high energies, especially for heavy lepton type neutrinos whose mean energy reaches $\sim90$\,MeV, are observed during the infall phase of PNS envelope and (2) a steady state neutrino emission of electron type neutrinos in the BH accretion phase.
Possible observations of high energy neutrinos from BH formation are also reported in a previous similar (but spherical symmetric) study by \cite{Gullin&O'Connor22}.
We attribute the first feature to more efficient isoenergy scatterings between neutrinos, which strive to emerge from the shock surface, and infalling stellar mantle ahead of the shock, which is mainly composed of heavy nuclei.
Using time evolution of neutrino spectral property, we showed that propagation of high energy neutrinos is indeed hindered, when the PNS shock surface drastically collapses (i.e. $1$\,ms$\lesssim t_{\rm BH}\lesssim2$\,ms).
Once the shock surface is engulfed by the BH, those neutrinos are radiated away, with some time delays for higher energy neutrinos.
In the BH accretion phase, the main component of accretion flow is high-$Y_e$ stellar mantle, whose temperature is at the highest a few MeV.
Therefore the main neutrino emission channel is the electron capture on heavy nuclei occurring in the vicinity of AH.
It results in a nearly constant electron type neutrino luminosity as also reported in \cite{Sykes&BMuller23}.
We would like to emphasize that these neutrino properties could be revealed only by full neutrino radiation-hydrodynamic simulations with numerical relativity without excising the relevant region outside the AH, i.e., by fully solving the region outside the BH.

In this study we employed only one non-rotating progenitor model.
In our future works, we are interested in exploring various CCSN models accompanied by BH formation.
For instance, a fallback scenario is one of the interesting topics.
The current progenitor model has a significantly high compactness $\xi_{2.5}=1.0$ at precollapse stage (\cite{O'Connor11} and see also Table\,1 in \cite{KurodaT22}), which leads to strong mass accretions during the PNS contraction phase.
Therefore it induces the PNS core-collapse without affording an opportunity for shock revival.
However, if one considers less compact stars \citep{Chan&Muller18,Powell21} or rotating stars \citep{Rahman22}, the shock revival aided by neutrino heating could happen before BH formation.
Such systems could be observed as a faint supernova \citep{Kochanek08,Adams17} and should be distinguished from the current failed SN (or direct BH formation) model with no shock revival.
Progenitor model dependency should definitely be explored in the future study to explain various observations.

Another interesting topic to be explored is the collapsar scenario \citep{MacFadyen99} as a possible route to long gamma-ray bursts and hypernovae.
In the collapsar scenario, a BH surrounded by a massive disk is formed, i.e., highly non spherical system is formed. Such systems can be followed only in numerical relativity with no approximation like CFC approximation.
For instance, after the formation of a massive disk, viscous effects significantly heat the disk, leading eventually to the launch of energetic outflows \citep[in the context of both NS mergers and massive stellar collapse, see, e.g.,][]{Fernandez13,Just15,Fujibayashi2020a,Fujibayashi2020b,Fujibayashi23}.
As another intriguing and also a challenging topic in the context of collapsar scenario, the impact of magnetic fields threading the central BH is undoubtedly worth to be explored as a possible origin of relativistic jets generated via, e.g., the Blandford-Znajek mechanism \citep{Blandford77}.
It has been recently demonstrated by \cite{Christie2019dec,Hayashi22} that the Blandford-Znajek mechanism is a promising mechanism for launching a jet, but only in the framework of compact mergers.
We will explore this fascinating topic in our future CCSN studies.

\section*{Acknowledgements}
We acknowledge K. Kiuchi, S. Fujibayashi, and A. Betranhandy for fruitful discussions.
This work was in part supported by Grant-in-Aid for Scientific Research (Nos. 20H00158 and 23H04900) of Japanese MEXT/JSPS.
Numerical computations were carried out on Sakura and Raven clusters at Max Planck Computing and Data Facility.

\section*{Data Availability}
The data underlying this article will be shared on reasonable request to the corresponding author.



\bibliographystyle{mnras}
\bibliography{mybib} 








\bsp	
\label{lastpage}
\end{document}